\documentclass[aps,pre,preprint,groupedaddress,showpacs]{revtex4-1}
\usepackage[dvips]{graphicx}
\usepackage{amssymb}
\usepackage{bbold}

\newcommand{\mc}{\multicolumn}

\begin{document}

\title{
\Large\bf Exploiting the hopping parameter expansion in the hybrid Monte Carlo
(HMC) simulation of lattice QCD with two degenerate flavours of Wilson fermions}

\author{Martin Hasenbusch}
\email[]{Martin.Hasenbusch@physik.hu-berlin.de}
\affiliation{
Institut f\"ur Physik, Humboldt-Universit\"at zu Berlin,
Newtonstr. 15, 12489 Berlin, Germany}

\date{\today}

\begin{abstract}
We show how the hopping parameter expansion at order $\kappa^2$ and 
 $\kappa^4$ can be exploited in the simulation of lattice QCD with 
two flavours of degenerate Wilson fermions. A natural extension of this 
idea is a ``UV-filtering'' by using rooted polynomials. These approaches 
can be easily combined with, for example, mass preconditioning.
First numerical tests are performed for the Wilson gauge action at
$\beta=5.6$ and $\kappa=0.156$ and $0.1575$.
\end{abstract}

\maketitle

\section{Introduction}
Lattice QCD simulations are our primary tool to obtain non-perturbative 
results from QCD.  To this end, a fair share of CPU time on the largest 
supercomputers that are available today is used. Still we would not mind,
getting more accurate results from such simulations. Hence any algorithmic 
progress is highly desirable.
Here I shall address the generation of the gauge field.  

In order to fix the notation, let us briefly recall the definition of lattice 
QCD. It is
defined on a four dimensional hypercubic lattice. On the links of the
lattice there are matrices $U_{x,\mu} \in \mbox{SU(3)}$, representing
the gluon field, where $x$ denotes a site of the lattice and $\mu \in \{0,1,2,3\}$
labels the directions. The fields that represent the fermions live on the
sites. These fields assume Grassmanian values. The interactions are
encoded by the Euclidian action
\begin{equation}
 S[U,\psi,\bar \psi] = S_G[U] + S_F[U,\psi,\bar \psi] \;.
\end{equation}
The fermion fields $\psi$, $\bar \psi$ appear in bilinear form, and
therefore can be integrated out exactly in the partition function $Z$.
It remains an integral over the gauge field only:
\begin{equation}
\label{partition}
 Z = \int \mbox{D}[U] \exp(-S_G[U])  \;\; \prod_{f=1}^n \mbox{det} M_f[U] \;\; ,
\end{equation}
where $M_f[U]$ is the fermion matrix and the product runs over the flavours
of the quarks.
In the literature  different types of fermion actions are
discussed. In the following we shall consider two degenerate flavours
of  Wilson fermions.  The fermion matrix is given by 
\begin{equation}
 M = \mathbb{1}  - \kappa  H \;\;, 
\end{equation}
where
\begin{equation}
H_{xy} =  \sum_{\mu} \left\{
   (1-\gamma_{\mu} ) \; U_{x,\mu} \; \delta_{x+\hat \mu,y} \; +  \;
   (1+\gamma_{\mu} ) \; U_{x-\hat \mu,\mu}^{\dag} \; \delta_{x-\hat \mu,y} \right\} \;\;,
\end{equation}
is the hopping matrix and the hopping parameter $\kappa$ is a real number. 
The Wilson plaquette action is given by
\begin{equation}
S_G[U] = - \frac{\beta}{3} \sum_{x} \sum_{\mu>\nu}
\mbox{Re} \; \mbox{Tr} \;
\left ( U_{x,\mu} U_{x+\hat \mu,\nu} U_{x+\hat \nu,\mu}^{\dag}
U_{x,\nu}^{\dag} \right ) \;\;,
\end{equation}
where $\hat \mu$ is a unit vector in $\mu$-direction. For a more 
detailed discussion see for example the textbooks and review articles
\cite{Rothe,review,review2,Gupta97}. 

For lattice sizes that are needed to extract continuum physics from
the simulation, it is by far too expensive to evaluate the determinant 
of the fermion matrix exactly. Therefore, 
following the proposal of Weingarten and Petcher \cite{WePe},
in the case of two degenerate flavours, one introduces  auxiliary 
degrees of freedom, so called pseudo-fermions:
\begin{equation}
\label{Fintegral}
\mbox{det} M^\dag M \; \propto \; \int \mbox{D}[\phi] \mbox{D}[\phi^\dag] \;
\exp(-S_{PF}) \;\;,
\end{equation}
where
\begin{equation}
\label{WePe_action}
 S_{PF} = |M^{-1} \phi|^2  \;\;,
\end{equation}
where $ \phi$ is a vector with complex components. 
Hence the action, as a function of the gauge field and the
pseudo-fermion fields, is given by
\begin{equation}
\label{WePe_action2}
S(U,\phi)\; = \;S_G(U)\; +\; S_{PF}(U,\phi)  \;\;. 
\end{equation}
Still the pseudo-fermion action is non-local and the evaluation requires 
the solution of a system of linear equations. The non-locality is in particular a problem for local algorithms that are used to simulate the pure gauge 
action. The hybrid Monte Carlo (HMC) algorithm  \cite{HMC} is better adapted
to this situation, since all gauge degrees of freedom evolve simultaneously. 
To this end, an artificial Hamiltonian is introduced:
\begin{equation}
 H = \frac{1}{2} \left(\Pi,\Pi \right) + S(U,\phi) \;, 
\end{equation}
where the antihermitian  momenta $\Pi_{x,\mu}$ are conjugate to the gauge 
field $U_{x,\mu}$.  They are auxiliary variables that are solely introduced
for algorithmic reason.  Their scalar product is defined as
\begin{equation}
\label{Sproduct}
 \left(\Pi,\Pi \right) = -2 \sum_{x,\mu} \mbox{Tr} \Pi^2_{x,\mu} \;. 
\end{equation}
Here we follow the convention of, for example, ref. \cite{domain2}.  
Note that in the literature often the factor $2$ is omitted in the definition
of the scalar product, see e.g. ref. \cite{Gupta89}.
Note that this leads to a relative factor $\sqrt{2}$ in the fictitious 
Monte Carlo time $\tau$ that is introduced below.
A discussion of this point is given in ref. \cite{Schaefer}, below eq.~(3.2).
The momenta and the gauge field evolve according the equations of motion
\begin{equation}
\label{Uevolution}
 \frac{\mbox{d} }{\mbox{d} \tau}  \Pi_{x,\mu} = - { \cal F}_{x,\mu}
\;\; \mbox{and} \;\;  \frac{\mbox{d} }{\mbox{d} \tau} U_{x,\mu} =
 \Pi_{x,\mu}  U_{x,\mu} \;,
\end{equation}
where $\tau$ is the fictitious Monte Carlo time and
the force ${\cal F}$ fulfills $(\omega, {\cal F}) = \delta_{\omega} S(U)$ for
infinitesimal variations of the gauge field
$\delta_{\omega} U_{x,\mu} = \omega_{x,\mu} U_{x,\mu}$. Here we consider 
the so called  $\phi$-algorithm \cite{phialg}, where the pseudo-fermions 
stay fixed during the evolution of the gauge field and the momenta.

The equations of motion~(\ref{Uevolution}) can not be integrated exactly. 
Therefore a numerical integration scheme with a finite step-size is used.
This leads to an integration error. The idea of the HMC-algorithm 
\cite{HMC} is that this error can be corrected for by a Metropolis  
accept/reject step.

One update cycle (or trajectory) of the HMC is composed of the following 
three steps:
\begin{itemize}
\item
Perform a heat-bath for both the conjugate momenta $\Pi$ and the 
pseudo-fermion field $\phi$. 
In the case of the pseudo-fermion field one 
generates a field $\eta$ with a Gaussian distribution 
$P(\eta) \propto \exp(- |\eta|^2)$  and then 
\begin{equation}
\label{initPseudo}
\phi = M \eta \;\;.
\end{equation}
Evaluate the Hamiltonian 
\begin{equation}
 H(U,\phi,\Pi) = S_G(U) + |\eta|^2 + \frac{1}{2} \left(\Pi,\Pi \right) 
\end{equation}
and save the initial gauge configuration $U$.

\item
Keeping $\phi$  fixed, we evolve the gauge field $U$ and the conjugate momenta
$\Pi$ according to the classical equations of motion for the fictitious time
$\tau$. Since this can not be done exactly, a numerical integration scheme
with the finite step-size $\delta \tau$ is used. At the end of the integration
we have the fields $U'$, $\Pi'$, and $\phi'=\phi$.  For a detailed discussion 
of the integration scheme see below.

\item
Accept $U'$ as the new gauge field with the probability
\begin{equation}
 P_{acc} = \mbox{min}[1,\exp(-\Delta H)]  \;\; , 
\end{equation}
where 
\begin{equation}
 \Delta H =  H(U',\phi,\Pi') -  H(U,\phi,\Pi)
\end{equation}
else we keep $U$. 

\end{itemize}

In order to fulfill detailed balance, the numerical integration scheme has 
to be area preserving and reversible. Reversible means that 
changing the sign of the momenta at the end of the integration time, we 
run back exactly to the initial gauge field $U$. Such integration schemes 
are called symplectic integrators.
Let us introduce a short hand for finite update steps  by $\delta \tau$:
\begin{eqnarray}
P(\delta \tau): \; \Pi_{x,\mu} \rightarrow  \Pi_{x,\mu}'  &=& \Pi_{x,\mu} + \delta \tau {\cal F}_{x,\mu} \;, \\
T(\delta \tau): U_{x,\mu} \rightarrow   U_{x,\mu}'   &=&  \exp(i \delta \tau \Pi_{x,\mu}) \; U_{x,\mu} \;.
\end{eqnarray}
Itegrators are build from these basic steps.
Here we consider the second order Omelyan integrator \cite{Omel03}, 
\begin{equation}
\label{Omel}
 T_{O} = P(\lambda \delta \tau) \;
              T(\delta \tau/2) \;
               P([1-2 \lambda] \delta \tau) \;
              T(\delta \tau/2) \;
           P(\lambda \delta \tau)  \;\;, 
\end{equation} 
where we get for $\lambda=1/6$ the scheme proposed in ref. \cite{Sex92},
which is also discussed for example in ref. \cite{Bussone18}.  
A trajectory of length
$\tau$ is given by $T_{O}^m$ with $\tau=m \; \delta \tau$. Taking $\lambda=1/2$
the expression~(\ref{Omel}) simplifies to the well know leapfrog scheme:
\begin{equation}
 T_{L} = P(\delta \tau/2) \;
              T(\delta \tau) \;
           P(\delta \tau/2)  \;.
\end{equation}
In our simulations,  we use both the leapfrog and the Omelyan scheme 
with $\lambda=1/6$.
Sexton and Weingarten proposed a multilevel integration scheme \cite{Sex92}. 
Each level is associated with a term in the action. For example,  
in eq.~(\ref{WePe_action2}), we can
associate the gauge action with level $i=0$ and the pseudo-fermion action 
with $i=1$. For each level a time step 
$\delta \tau_i=2 m_{i-1} \delta \tau_{i-1}$ is defined.
The scheme can be iteratively defined:
\begin{equation}
T_{SW,i} =P_i(\lambda \delta \tau_i) \; \left[T_{SW,i-1}  \right]^{m_{i-1}} \; 
          P_i([1-2 \lambda] \delta \tau_i) \;
          \left[T_{SW,i-1}  \right]^{m_{i-1}} \; P_i(\lambda \delta \tau_i)
\end{equation}
and
\begin{equation} 
T_{SW,0} =  P_0(\lambda \delta \tau_0) \;
              T(\delta \tau_0/2) \;
               P_0([1-2 \lambda] \delta \tau_0) \;
              T(\delta \tau_0/2) \;
           P_0(\lambda \delta \tau_0)  \;\; \;. 
\end{equation}
Note that for the leapfrog scheme, we use the convention  
$\delta \tau_i= m_{i-1} \delta \tau_{i-1}$, which is more natural in this 
case.
For a nice discussion of this scheme see for example section 2.2 of 
ref. \cite{urbach}.  The scheme can be generalized even further. The parameter
$\lambda$ might depend on the level $i$. Or me might use a fourth order 
scheme at low levels and a second order scheme at higher levels.
An important property of symplectic  integrators is that they preserve 
a so called shadow Hamiltonian. Here we will not delve into this 
discussion but refer the reader to refs. \cite{shadow,Bussone18} and 
references therein.

Applying the HMC algorithm to the pseudo-fermion action~(\ref{WePe_action}),
two problems are encountered: Going to lighter  quark masses, sending 
$\kappa$ to $\kappa_c$, the condition number of the fermion matrix increases.
As a result, for iterative solvers like the Conjugate Gradient (CG) or the 
Biconjugate gradient stabilized method
(BiCGstab) \cite{BiCG1,BiCG2},  the number of iterations needed to solve the 
system of linear equations is increasing.  The second problem 
is less obvious. It turns out that, in order to keep the acceptance rate fixed,
the step-size of the integration scheme has to be reduced with decreasing 
quark mass.
At the Lattice 2001 in Berlin the
situation was referred to as ``Berlin wall''.  At the time, it seemed 
impossible to reach sufficiently small masses, to reliably extrapolate, by
using chiral perturbation theory, to the physical mass of the pion.

The situation considerably improved by the advent of better solvers, for example
\cite{Luescherdefl,Multigrid}, and by replacing the pseudo-fermion
action~(\ref{WePe_action}) by better alternatives.
Note that the representation of the fermion determinant by
pseudo-fermions is not unique.  In \cite{multiboson}  a large number 
of pseudo-fermion fields were introduced, allowing to express the fermion 
determinant in terms of a local pseudo-fermion action. This approach 
did not outperform the HMC algorithm in the end. It turned out that the 
large number of fields implicate that only small steps can be performed in the 
update. An alternative approach to local updating, which also did not
outperform the HMC, is discussed in ref. \cite{Myfinite}.
See also \cite{Knechtli13} and references therein.

Based on this experience, alternatives to eq.~(\ref{WePe_action}),
to be used in HMC simulations, were proposed.
 These are primarily mass preconditioning
\cite{MyHasenbusch,masspreconditioning2}, domain-decomposition 
\cite{domain,domain2},
and rooting \cite{Clark007}.  The basic idea behind these approaches is 
to split the fermion matrix $M$ into (several) factors, and introduce 
a separate pseudo-fermion field for each of the factors.  By using a suitable
factorisation, the stochastic estimate of the fermion determinant becomes
less noisy, allowing for a larger step-size in the integration scheme.
A second potential advantage is that different parts of the pseudo-fermion 
action can be put on different time-scales of the 
integration scheme \cite{Sex92}.
In the ideal case, the numerically most expensive parts can be put on 
large time scales.

In the case of a finite step updating scheme \cite{Myfinite}, 
the multiboson (MB) algorithm \cite{multiboson} and the polynomial hybrid Monte
Carlo (PHMC) algorithm \cite{ForcrandPHMC,PHMC,PHMC2}, it has been shown 
that the 
updating scheme becomes more efficient by incorporating the hopping
parameter expansion \cite{ForcrandUV,Forcrand,Ishikawa:2006pb}. The hopping 
parameter expansion, taken at a low order, is used as UV-filter for the 
pseudo-fermion action.
Here we demonstrate how this can efficiently be done for the 
HMC algorithm applied to two degenerate flavours. Compared with the 
simulation using the pseudo-fermion action~(\ref{WePe_action}) we get a
speed-up of a factor of two or three, depending on the order of the 
hopping parameter expansion.
In large scale
simulations, this idea can be combined with mass preconditioning 
\cite{MyHasenbusch} and might lead to a speed-up of the order of 
$20 \%$. Furthermore we give a preliminary discussion of UV-filtering 
by using rooted polynomials. The motivation is similar to ref. 
\cite{Clark007} and could also be seen as a natural extension of 
the UV-filtering by using the hopping parameter expansion.

The outline of the paper is the following. In the next section we discuss
in detail how the hopping parameter expansion is used as UV-filter.
Then we discuss how this idea can be naturally extended by using  
polynomial approximations of the rooted inverse of the fermion matrix. 
We briefly summarize results on the acceptance rate, the variance 
of $\Delta H$ and the forces that are given in the literature.
Then in section \ref{NumRes} we discuss our numerical results. 
First we study the effect of UV-filtering by using the hopping parameter
expansion up to the orders $\kappa^2$ and $\kappa^4$. Then  we present our
still very preliminary results on rooted polynomials.
Finally we give a summary and an outlook.

\section{Incorporating the hopping parameter expansion into the 
hybrid Monte Carlo simulation}
In the case of two degenerate flavours, the fermion determinant can be 
expressed as 
\begin{equation}
\label{traceM}
\mbox{det} M^{\dag} M =  \exp(\mbox{Tr} \ln M^{\dag} + 
\mbox{Tr} \ln M) \;, 
\end{equation}
where one expands
\begin{equation}
\ln M =  \ln(1 - \kappa H) = - \sum_{n=1}^{\infty} \frac{1}{n} \kappa^n H^n \;.
\end{equation}
For small values of $n$,  $\mbox{Tr} H^n$ can be evaluated analytically.
In the case of Wilson fermions, terms with odd values of $n$ do not 
contribute. Furthermore  $n=2$ also does not contribute.
The leading non-vanishing 
contribution $\mbox{Tr} H^4$ amounts to a plaquette term. 
This can be written as a shift of the parameter $\beta$. In the case
of two degenerate Wilson fermions one gets $\Delta \beta= 96 \kappa^4$. 
For $n=6$ we get
contributions from three different Wilson loops. With increasing $n$, the 
number of Wilson loops that contribute, rapidly increase and things become 
intractable. For a more detailed discussion see sect. III  of 
ref. \cite{thron97}.
In the case of clover-improved Wilson fermions  the situation is
worse. There is already a non-vanishing contribution for $n=2$, see eq.~(2.8)
of ref. \cite{Ishikawa:2006pb}. Already $n=4$ was not considered in 
ref. \cite{Ishikawa:2006pb}, since it is too involved.

In the simulation we consider a modified gauge action
\begin{equation}
\label{ModGauge}
 \tilde S_G =  S_G  +2 \sum_{n=1}^{k} \frac{1}{n} \kappa^n \; \mbox{Tr} H^n \;,
\end{equation}
where $k$ is the order, up to which $\mbox{Tr} H^n$  in terms 
of Wilson loops is tractable in the simulation.

In ref. \cite{Myfinite} we discussed preconditioning by using the hopping 
parameter expansion in the context of a finite step updating scheme. To this
end the value of the pseudo-fermion action has to be evaluated.
Following eq.~(8) of ref. \cite{Myfinite} a modified fermion matrix
is introduced by
\begin{equation}
\label{tildeM}
\tilde M = M \; \exp\left(\sum_{n=1}^k \frac{1}{n} \kappa^n H^n \right) \; ,
\end{equation}
and correspondingly
\begin{equation}
\label{ModPF}
\tilde S_{PF} = |\tilde M^{-1} \phi|^2 \;\;.
\end{equation}
The idea is that $\tilde S_{PF}$ fluctuates less than $S_{PF}$ and hence
allows for a larger step-size in the HMC-simulation.
In ref. \cite{Myfinite} we evaluated $\tilde M^{-1} \phi$ by using the series
expansion of $\tilde M^{-1}$ in $ \kappa H$. Also in the case of the 
MB algorithm \cite{ForcrandUV,Forcrand} and the PHMC algorithm 
\cite{Ishikawa:2006pb}
it is natural to represent  $\tilde M$ by using a polynomial in $M$.

Here we discuss an alternative representation that is more suitable for
the HMC algorithm applied to two degenerate fermion flavours. In particular, 
we express $\tilde M^{-1}$ essentially in terms of $M^{-1}$ to make 
use of iterative solvers to compute $\tilde M^{-1} \phi$.
For simplicity, let us first discuss the case $k=1$.  
The series expansion of the inverse of $\tilde M$ in $\kappa H$ is given by
\begin{equation}
 \tilde M^{-1} = \exp(- \kappa H) (1-\kappa H)^{-1} 
= \sum_{n=0}^{\infty} 
a_n \kappa^n H^n  \;\;.
\end{equation}
Since all coefficients of the expansion of $M^{-1}$ are equal to one, 
we can easily evaluate the coefficients 
\begin{equation}
 a_n = \sum_{i=0}^{n} (-1)^{i} \frac{1}{i!} \;\;\; \;\;\; \;\;\; , \;\; \;\;\; 
\lim_{n \rightarrow \infty} a_n = \exp(-1) \;\;. 
\end{equation}
Hence we can write 
\begin{equation}
 \tilde M^{-1} =  \sum_{n=0}^{\infty} b_n \kappa^n H^n   +  \alpha M^{-1} \;,
\end{equation}
where $\alpha =  \exp(-1)$ and 
\begin{equation}
 b_n = - \sum_{i=n+1}^{\infty} (-1)^{i} \frac{1}{i!} \;. 
\end{equation} 
Since $b_n$ rapidly converges to $0$, the sum
\begin{equation}
\label{bSum}
 \sum_{n=0}^{\infty} b_n \kappa^n H^n
\end{equation}
can be truncated at a low order $n_{max}$. 
For larger values of $k$ we get a similar 
result, where $\alpha=\exp(-\sum_{n=1}^k  1/n)$. The coefficients $b_n$ can 
be evaluated by using an algebra program like Maple or Mathematica. 
Note that the coefficients in eq.~(\ref{ModGauge}) are tunable parameters 
of the algorithm.
Previous experience \cite{Myfinite,Ishikawa:2006pb} however shows that taking
the values given by the hopping parameter expansion is a good choice. Here we 
will not further discuss this question.

Now let us discuss how the HMC algorithm can be implemented for
$\tilde M^{-1}$.  The crucial question is how the forces can be 
computed. Here we have to put together the results obtained for the 
HMC algorithm and the PHMC algorithm. 
The variation of the pseudo-fermion action $S_{PF}$ with 
respect to the gauge field can be computed as 
\begin{equation}
 \delta S_{PF} = - X^{\dag}  \delta M Y  + h.c.  \;\;,
\end{equation}
where 
\begin{equation} 
X = (M M^{\dag})^{-1} \phi \;\;, \;\; \;\;  Y = M^{-1} \phi \;\;.
\end{equation}
The variation of the polynomial
has been worked out in ref. \cite{PHMC,PHMC2}. We follow the 
implementation of ref. \cite{PHMC2,Ishikawa:2006pb} using Horners scheme.
Here we need the variation of $\tilde S_{PF} = |\tilde M^{-1} \phi|^2$ with
\begin{equation} 
 \tilde M^{-1} = \alpha M^{-1} + \sum_{n=0}^{n_t} b_n \kappa^n H^{n} \;\;. 
\end{equation}
Note that we are free to take $n_t < n_{max}$, since the truncation error 
introduced is  corrected for in the accept/reject step, where the 
summation is performed up to $n_{max}$.  We get
\begin{equation}
\delta \tilde S_{PF} = \phi^{\dag} \tilde M^{-1 \; \dag} 
                     \delta \tilde M^{-1} \phi + h.c. \;,
\end{equation} 
where 
\begin{equation}
 \delta \tilde M^{-1} \phi = \left [- \alpha M^{-1} \delta M  M^{-1}
            + \sum_{n=1}^{n_t} b_n \kappa^n \delta (H^n)   \right] \phi \;,
\end{equation}
where
\begin{equation}
    \delta (H^n)  = \sum_{i=1}^{n} H^{i-1} \; \delta H \; H^{n-i} \;.
\end{equation}
In order to compute the variation for the polynomial efficiently, 
$n_t$ vectors have to be precomputed, following Horners scheme:
\begin{equation}
 Y_{n_t} = b_{n_t} \phi
\end{equation}
and then recursively 
\begin{equation}
 Y_{i-1} = b_{i-1} + \kappa H Y_{i} 
\end{equation}
down to 
\begin{equation}
Y_0 = \left[\sum_{i=0}^{n_t} b_i \kappa^i H^i \right] \phi \;.
\end{equation}
Then we compute recursively 
\begin{equation}
X_1 = Y_0 + \tilde Y \;\;\;, \;\; \mbox{where} \;\; \tilde Y = \alpha M^{-1} \phi
\end{equation}
and 
\begin{equation}
X_i = \kappa H X_{i-1} \;. 
\end{equation}
The variation of the pseudo-fermion action can be written as
\begin{equation}
\delta \tilde S_{PF} = \kappa \; \tilde X^{\dag} \delta H \tilde Y \;+ \;
\kappa \sum_{i=1}^{N} X_i^{\dag} \delta H Y_i  \;\; + \;\; h.c. \;,
\end{equation}
where $\tilde X = M^{-1} X_1$. 

In the following we refer to exploiting the hopping parameter expansion up
to order $\kappa^k$ as $\kappa^k$-filtering.

\subsection{Rooted polynomials}
In our simulations we make use of the hopping parameter expansion up 
to $\kappa^4$. It is practically impossible to push the hopping parameter 
expansion to higher order. Therefore, with a similar motivation as ref.
\cite{Clark007}, where the rational HMC is considered,  
we propose to use rooted polynomials as UV-filters. Also note that 
\begin{equation}
 \lim_{N \rightarrow \infty}  
 \exp \left(- \sum_{i=1}^N  |M^{-1/N} \phi_i|^2  \right) \propto 
 \mbox{det} M^{\dag} M \;\;.
\end{equation}
For a discussion see section II. B. of ref. \cite{Myfinite}. This means 
that for sufficiently large $N$, we can approximate the hopping parameter
expansion by using low order polynomials that approximate $M^{-1/N}$. 
Let us define $M_0 = \tilde M$, eq.~(\ref{tildeM}), 
and then recursively
\begin{equation}
 M_{j+1}  = W_j^{-N_j} M_j
\end{equation}
up to some maximal $j_{max}$, where 
\begin{equation}
 W_j^{-1}  = \sum_{i=0}^{n_j} a_{j,i} \kappa^i H^i 
= M_j^{-1/N_j} + \mbox{O}\left(\kappa^{n_j+1}\right) \;\;,
\end{equation}
where $n_j > n_{j-1}$. The remainder can be written as
\begin{equation}
\label{restM}
 M_{j_{max}+1}^{-1}  = 
\sum_{n=0}^{\infty} b_n \kappa^n H^n + \alpha M^{-1} \;\;, 
\end{equation}
where $b_n$ and $\alpha$ are computed by using an algebra program. 

The construction proposed here contains both the noise reduction by 
rooting as proposed in \cite{Myfinite,Clark007} as well as a hierarchical 
splitting similar to mass preconditioning.  Note that a hierarchical splitting, 
in the framework of the PHMC, was already discussed in refs. 
\cite{Kamleh12,Kamleh17,Kamleh18}.

In particular, aiming at the application to a single flavour, one
would like to investigate how well $M_{j_{max}+1}^{-1/N}$ can be approximated
by a rational approximation. Also it might be feasible to 
compute $\mbox{det} M_{j_{max}+1}$, without using a noisy estimator, since 
likely only a few smallest eigenvalues of $M$  contribute. In this case, 
it might be sufficient to compute $\mbox{det} M_{j_{max}+1}$ in the 
accept/reject step only.

In our numerical tests we have used $j_{max}=2$ and $N_1=N_2=N$ for
simplicity. The general framework contains a large number of free 
parameters that is hard to tune without having a theoretical understanding
of the dependence of the acceptance rate on these parameters. Ref. 
\cite{Bussone18} and possible extensions might be helpful to this end.

In the case of the pseudo-fermion action~(\ref{WePe_action}) it is 
simple to perform a heat-bath update, eq.~(\ref{initPseudo}), of the  
pseudo-fermions at the beginning of the trajectory.  The fermion 
matrix $M$ has to be applied to a vector with a Gaussian distribution.
In the case of the rooted polynomials the numerical costs are 
considerable larger, since $W_{j}$
has to be represented by a high order polynomial in $M$ or equivalently $H$. 
In our preliminary study, we implemented the heat-bath update of the 
pseudo-fermions associated with $W_{j}$ in the straight forward way. 
A more efficient solution is provided by ref. \cite{Forcrand99}, 
where only a good approximation of $W_{j}$ is needed to update the 
pseudo-fermions. 

\subsection{Even/odd preconditioning}
In all our numerical tests, we started from the even/odd preconditioned  
fermion matrix
\begin{equation} 
 M_{oo} = \mathbb{1}_{oo}  - \kappa^2 H_{oe}  H_{eo} \;\;,
\end{equation}  
where $e$ and $o$ denote the collection of even and odd sites, respectively.
Note that $\mbox{det}  M_{oo} = \mbox{det} M$ and the condition number of
$M_{oo}$ is reduced compared with $M$. In the discussion of the algorithm
above, essentially $\kappa H$ has to be replaced by $\kappa^2 H_{oe}  H_{eo}$.
Note that indices in section \ref{NumRes} below, refer to powers of 
$\kappa^2 H_{oe}  H_{eo}$.
Note that in ref. \cite{PHMC2} it is explicitly spelled out, how the PHMC 
algorithm can be implemented for even/odd preconditioned clover-improved
Wilson fermions.

\section{The acceptance rate and forces}
Typically the step-size of the HMC is tuned such that the acceptance 
rate $0.8 \lessapprox P_{acc} \lessapprox 0.9$.  The optimal value depends 
on the integration scheme that is used. Also the occurrence of spikes might
require to decrease the step-size $\delta \tau$.  Spikes mean that 
occasionally  $\Delta H \gg 1$ appears in the simulation.  Here we have 
encountered this phenomenon when using the second order Omelyan integrator.

The acceptance rate can be determined by simply counting the accepted 
configuration. The statistical error is reduced by sampling 
$\mbox{min} [1,\exp(-\Delta H)]$ instead.
Detailed balance implies 
\begin{equation}
\label{detailed}
\langle \exp(-\Delta H) \rangle = 1  \; . 
\end{equation}
It is a useful check for the correctness of the program to sample 
$\exp(-\Delta H)$ and check whether the average is consistent with one.
Based on eq.~(\ref{detailed}) one can derive for high acceptance rates
\begin{equation}
\label{creutz}
P_{acc} = \mbox{erfc} \left( \sqrt{\mbox{Var}(\Delta H)/8} \right) \;\;.
\end{equation}
See eq.~(3.1) of ref. \cite{Bussone18} and references therein. In our 
simulations, as long as no spikes occur, eq.~(\ref{creutz}) turned out
to be valid to good precision.

The HMC simulation using improved pseudo-fermion actions \cite{masspreconditioning2,domain2,Clark007,Kamleh12,Kamleh17,Kamleh18}
requires to tune a number of parameters. Therefore it is highly desirable
to know how the acceptance rate, or equivalently Var$(\Delta H)$, depends
on these parameters.
A step in this direction is taken by ref. \cite{Bussone18}, where the 
variances of the forces associated with the different parts of the action 
are related to Var$(\Delta H)$.
For the second order Omelyan scheme with $\lambda=1/6$ the authors of ref.
\cite{Bussone18} find, see their eq.~(3.4),  
\begin{equation}
\label{masterF}
 \mbox{Var}(\Delta H) = \frac{2 \delta \tau^4}{72^2} 
\left[\mbox{Var}(|{\cal F}_{i_{max}}|^2) + 
\frac{\mbox{Var}(|{\cal F}_{i_{max}-1}|^2)}{ (4 m_{i_{max}-1}^2)^2} + ... \right]
\end{equation}
Note that for 
$\lambda \ne 1/6$ also other terms than the forces appear at the order
$\delta \tau^4$. For a more general result see ref. \cite{Thesis}. 
A main ingredient in the derivation of eq.~(\ref{masterF}) 
is the fact that a symplectic integrator conserves a shadow Hamiltonian.
The deviation of the shadow Hamiltonian from the true Hamiltonian can be 
computed as a power series in the step-size $\delta \tau$.  Furthermore, 
it is assumed that the forces due to different pieces of the action are 
not correlated.

\section{Numerical results}
\label{NumRes}
The study is performed on three servers with two CPUs with 10 cores each, that 
were immediately available to us. For programming convenience no
highly optimized code was used.  As solver, we have used the 
BiCG-stab \cite{BiCG1,BiCG2} algorithm. 
Here we did not experiment much with the 
stopping criterion, but did run the solver essentially up to machine 
precision.
We simulate comparatively small lattices at $\beta=5.6$.  
In particular we have tested $\kappa^2$- and $\kappa^4$-filtering 
extensively by simulating a $12^3 \times 24$ lattice at $\kappa=0.156$.
To consolidate the result, two simulations of a $16^3 \times 32$ lattice at
$\kappa=0.1575$ are performed. 
Our preliminary study of the performance of the HMC 
using rooted polynomials are also performed for a $16^3 \times 32$ lattice 
at $\kappa=0.1575$. The linear lattice sizes are measured in units of the 
lattice
spacing $a$. We use periodic boundary conditions in spacial direction.
In the case of the temporal direction, periodic boundary conditions are employed
for the gauge action and anti-periodic ones for the fermion action.

A rather detailed study at this value of $\beta$ is presented in 
ref. \cite{lippert}.  Based on the Sommer scale $r_0$ \cite{Sommer},
the authors of
ref. \cite{lippert} find that for $\beta=5.6$, $\kappa=0.156$ on a 
$16^3 \times 32$ lattice $a=0.09796(56)$ fm. For the same parameters they
find $m_{PS}=0.9002(69)$ GeV for the mass of the lightest pseudo-scalar 
particle. 
For $\beta=5.6$, $\kappa=0.1575$ on a $16^3 \times 32$ lattice they obtain
 $a=0.0839(11)$ fm and $m_{PS}=0.6524(86)$  GeV.  This means that the masses
are still quite large compared with the  mass of the pion 
$m_{\pi^0} \approx 135$ MeV. 
Note that a number of algorithmic studies were performed at  $\beta=5.6$,
the values of $\kappa$ and lattice sizes that were studied in ref. 
\cite{lippert}. See for example \cite{domain2,urbach,Kamleh12}. 

\subsection{Exploiting the hopping parameter expansion} 
In this set of simulations, we tested 
the efficiency of $\kappa^k$-filtering. To 
this end, we simulated the system with the pseudo-fermion 
action~(\ref{WePe_action}) and the modified  pseudo-fermion 
action~(\ref{ModPF}) up to $\kappa^2$ and $\kappa^4$. 
We simulated by using the leapfrog as well as the second order Omelyan
integrator at $\lambda=1/6$.  In both cases, we used two time scales. 
On the coarse time step we put the pseudo-fermion action and on the 
fine one the gauge action. The time step of the gauge action was chosen 
to be such that further decreasing it, virtually does not increase 
the acceptance rate. 
Next we have to decide how to truncate eq.~(\ref{bSum}). 
In the extended runs that we performed first, we set
ad hoc $n_t =7$  and $n_{max} = 19$ for $\kappa^2$. Note that
$b_7 = 1.98 ... \times 10^{-4}$ and $ b_{19} = 8.22 ... \times 10^{-18}$.
Instead, for  $\kappa^4$ we took $n_{t} =15$   and 
$n_{max} =29$, where $b_{15}= 7.91 ... \times 10^{-6}$ and
$b_{29}=1.15 ... \times 10^{-14}$. Later we carefully checked the 
dependence of the acceptance rate on $n_t$.  Furthermore we demonstrate that
the value of $n_t$ has no influence on the reversibility. 

\subsubsection{Extended runs}
We performed a few extended runs.
This way we checked for spikes in $\Delta H$ and tried to estimate
autocorrelation times.  Throughout we used trajectories of the length
$\tau= \sqrt{2}$, corresponding to $\tau=1$ in the convention of, for example,
ref. \cite{Gupta89}. 

A first set of runs was performed by using the leapfrog integration scheme.
We performed preliminary simulations to find the step-size $\delta \tau$
that gives $P_{acc} \approx 0.8$.  In table \ref{extendedleap} we summarize the
results of our extended runs.
The plaquette value is
$ P=\frac{1}{3 N_p} \sum_p \mbox{Re} \mbox{Tr}  U_p $,
where the sum runs over all plaquettes on the lattice and $U_p$ denotes the
ordered product of the gauge variables around the plaquette $p$ and $N_p$
is the number of plaquettes.
\begin{table}
\caption{\sl \label{extendedleap}
Extended runs using the leapfrog algorithm for a $12^3 \times 24$ lattice 
at $\beta=5.6$ and $\kappa=0.156$. We study the effect of $\kappa^n$-filtering.
stat gives the number of trajectories,  the number of coarse time steps 
$m$, and the expectation value of the plaquette $\langle P \rangle$.  
The acceptance rate is given by
$P_{acc} = \langle \mbox{min} [1, \exp(-\Delta H)]\rangle $.  In all three 
cases, the estimate of $P_{acc}$ obtained from the variance Var($\Delta H$),
by using eq.~(\ref{creutz}) is consistent
with the result given in column 5.
}
\begin{center}
\begin{tabular}{cccccc}
\hline
 n  & $m$ & stat& $\langle P\rangle$& $P_{acc}$ & Var($\Delta H$) \\
\hline
  0    &     42     &  2770 &0.56982(7) &  0.8006(43) &  0.2673(54) \\
  2    &     21     &  7050 &0.56991(6) &  0.7981(26) &  0.2643(43) \\
  4    &     16     &  7610 &0.56995(4) &  0.8106(24) &  0.2264(40) \\
\hline
\end{tabular}
\end{center}
\end{table}
Since the effort required for the evaluation of the polynomial~(\ref{bSum}) is
small compared with that for the iterative solver, the  performance
gain achieved by the $\kappa^k$-filtering is essentially given by the 
ratio of the step numbers $m$. This means that even in the case of 
$\kappa^2$-filtering that is still
achievable in the case of clover-improvement \cite{Ishikawa:2006pb}, 
we see a gain of a factor of two. Next we redid the exercise by using the 
second order Omelyan integrator.  In order to get an acceptance rate
of $\approx 80 \%$,  we find from preliminary simulations that $m=18$
and $8$ for the order $0$ and $2$ are needed, 
respectively. Hence the performance gain 
is even a bit larger than in the case of the leapfrog integrator. Performing
longer runs, spikes in $\Delta H$ appeared. Therefore we do not further 
discuss these runs. It is known that the second order Omelyan integrator is 
more susceptible to this problem than the leapfrog.  The problem can be cured
by reducing the step-size.
In the case of $\kappa^4$-filtering  
we could not find an $m$ that gives an acceptance rate of 
$\approx 80 \%$.  For $m=6$, the acceptance rate is considerably larger
and for $m=5$ it is smaller. 
We decided to perform a longer run for $m=6$.  From  24540 trajectories 
we get $\langle P \rangle =0.56991(2)$, $P_{acc}= 0.8830(15)$, 
and Var$(\Delta H)=0.0886(15)$. In this run no spikes appear.
We find that the direct determination of $P_{acc}$ and the 
result obtained from eq.~(\ref{creutz}) are consistent.   From this run 
we get the estimates $\tau_{int,P} =\sqrt{2} \times 9.3(1.7)$ and 
$\tau_{int,iter} = \sqrt{2} \times 24.8(4.5)$
for the integrated autocorrelation times of the plaquette and the iteration 
number of the solver, respectively. 
Given the relatively low accuracy of the autocorrelation
time, we are not able to decide whether the UV-filtering has an influence on 
the autocorrelation time.

\subsubsection{The forces}
As it is argued in ref. \cite{Bussone18}, the acceptance rate can be inferred 
from  the variance of the forces Var$(|{\cal F}|^2)$.  Computing 
Var$(|{\cal F}|^2)$ for $\kappa^4$-filtering, we get essentially consistent 
results 
from the run with the leapfrog and the second order Omelyan integrator.
We conclude Var$(|{\cal F}_{PF}|^2) = 57500(1000)$, 
where the error is only a
rough estimate. In the case of the runs without filtering and 
$\kappa^2$-filtering, using the leapfrog integration scheme, we get  
Var$(|{\cal F}_{PF}|^2) =11400000(200000)$ and $344000(4000)$, respectively.
The runs with the second order Omelyan scheme contain spikes in $\Delta H$.
These spikes can also be seen in ${\cal F}_{PF}$. As a result, 
Var$(|{\cal F}_{PF}|^2)$ is by far larger than for the runs with the leapfrog.
Excluding the spikes by hand, Var$(|{\cal F}_{PF}|^2)$ is much reduced, and 
very roughly consistent with what we find in the runs with the leapfrog
integrator.  
Following eq.~(\ref{masterF}), keeping 
Var$(|{\cal F}_{PF}|^2) \; \delta \tau^4$ 
fixed, should result in a fixed acceptance rate. Indeed,
$(11400000/344000)^{1/4} \approx 2.4$ and $(11400000/58000)^{1/4} \approx 3.7$
are roughly consistent with the speed-ups that we have observed directly. 

For the gauge action, we get from the runs with the leapfrog and the 
second order Omelyan scheme for both $\kappa^2$-filtering and no 
filtering consistent results that can be summarized as 
Var$(|{\cal F}_{G}|^2) = 28800000(400000)$. In the case of 
$\kappa^4$-filtering, due to the larger value of $\beta$ in $\tilde S_G$, 
we get  the larger value Var$(|{\cal F}_{G}|^2) =30000000(400000)$. 
We checked that, also according to eq.~(\ref{masterF}), our choices of 
$\delta \tau_G$ are small enough, not to influence the acceptance rate
markedly.

\subsubsection{Truncation of the series and reversibility}
In principle we can relax the accuracy of the calculation of the force 
to the point, where the acceptance rate is markedly 
affected. However it turned out that, using iterative solvers, the 
reversibility of the integration is increasingly violated with decreasing 
accuracy of the solution. With exact numerics, reversibility would be 
given at any 
precision of the solver. However we work with double precision numbers, 
and rounding errors occur.  Furthermore, iterative solvers approach the 
solution in a chaotic way. Hence, if we stop the solver at a moderate 
precision, deviations caused by rounding errors are blown up. 
This phenomenon does not occur when we evaluate a series with fixed 
coefficients. Therefore the truncation at the order $n_t < n_{max}$
of the sum~(\ref{bSum}) can be chosen such that the acceptance rate is 
reduced by little compared with larger values of $n_t$.  
We checked this reasoning for $\kappa^4$-filtering and the second
order Omelyan scheme at $\lambda=1/6$.
To this end, we selected ten configurations, which were separated by 400 
trajectories each from our extended run. For each of these configurations, 
we started a trajectory using the same parameters as for our extended run. 
At the end of the trajectory the momenta are 
reversed and the trajectory is run backwards, resulting in the configuration
$U'$.  We compute  
\begin{equation}
\label{reverse}
 \Delta = \sum_{x,\mu}  |U_{x,\mu} - U_{x,\mu}'|^2  \;\;.
\end{equation} 
For $n_t=5,6$, and $15$ and running the solver essentially up to machine 
precision, we get $\Delta \approx 6.3 \times 10^{-21}$ for all three
choices. Instead, keeping $n_t=15$ fixed and relaxing the stopping 
criterion of the BiCG-stab, $\Delta$ is clearly increasing.

\subsubsection{The acceptance rate as a function of $n_t$}
For both $\kappa^2$- and $\kappa^4$-filtering, we performed  runs 
with different values of $n_{t}$. We used the second order Omelyan scheme 
with $\lambda=1/6$ throughout.
In all cases the trajectories have the length $\tau=\sqrt{2}$.
As expected, we find that with increasing $n_{t}$  the acceptance rate
rapidly reaches a plateau  value.

For $\kappa^2$-filtering, we first performed runs with $m = 8$. Similar 
to the extended run, spikes appeared. Therefore we redid the runs with 
$m = 10$, where we did not encounter this problem for $n_t>3$. 
The results
are summarized in table \ref{acceptntk2}.  The acceptance rate as well 
as Var$(\Delta H)$ rapidly approach a plateau, which is reached at the
level of our numerical precision for $n_t = 5$.  For $n_t>3$, 
the estimate of $P_{acc}$ obtained from the variance Var($\Delta H$),
by using eq.~(\ref{creutz}) is consistent with the direct measurement.

\begin{table}
\caption{\sl \label{acceptntk2}
Numerical results for $\kappa^2$-filtering. Simulations are performed with the 
second order Omelyan scheme at $\lambda=1/6$.
We give the acceptance rate and Var$(\Delta H)$
as a function of the maximal summation index $n_t$.  
Throughout we use $m=10$ 
and the length of the trajectory is $\tau=\sqrt{2}$. In the run 
for $n_t=3$ spikes occurred. After removing them by hand we get 
 Var$(\Delta H)= 0.88(5)$.
}
\begin{center}
\begin{tabular}{rrll}
\hline
 $n_t$  & stat & $P_{acc}$ & Var$(\Delta H)$ \\
\hline
  3 & 1000 & 0.6420(11) & \mc{1}{c}{-}  \\
  4 & 2500 & 0.9187(21) & 0.0424(14) \\
  5 & 2000 & 0.9361(21) & 0.0268(9) \\
  6 & 1950 & 0.9412(20) & 0.0249(8) \\
  7 & 2000 & 0.9406(20) & 0.0242(9) \\
\hline
\end{tabular}
\end{center}
\end{table}

Our results for $\kappa^4$-filtering are summarized in table 
\ref{acceptntk4}.  Also
here the acceptance rate as well as Var$(\Delta H)$ rapidly reach a plateau
value. At the level of our accuracy this happens for $n_t=7$. As expected, 
this value is larger than for $\kappa^2$-filtering. 

\begin{table}
\caption{\sl \label{acceptntk4}
Same as table \ref{acceptntk2} but for $\kappa^4$-filtering. 
Here we use $m=6$ throughout.
}
\begin{center}
\begin{tabular}{rrll}
\hline
 $n_t$  & stat & $P_{acc}$ & Var$(\Delta H)$ \\
\hline
   3    & 200  & 0.22(3)   & 4.98(59) \\
   4    & 1030 & 0.177(8)  & 8.12(24) \\
   5    & 6400 & 0.8631(22)& 0.1180(31) \\
   6    & 2200 & 0.8506(45)& 0.1512(63) \\
   7    & 2200 & 0.8868(32)& 0.0920(36) \\
   8    & 2000 & 0.8845(31)& 0.0848(29) \\
   9    & 2200 & 0.8851(36)& 0.0904(33) \\
  15    & 24500& 0.8830(15)& 0.0886(15) \\
\hline
\end{tabular}
\end{center}
\end{table}

We conclude that the choice of $n_t$ is uncritical. Using a few short runs 
we can locate the point, where the acceptance rate as a function of $n_t$
levels off. In the production run we then  add a small safety margin.

\subsubsection{Scaling with the lattice size and $\kappa$}
To get an idea how the performance scales with the hopping parameter
$\kappa$, we performed two short runs at $\kappa=0.1575$ on a $16^3 \times 32$
lattice by using $\kappa^4$-filtering. In both cases the length of a 
trajectory is $\tau=\sqrt{2}$. We started
the simulations with a configuration taken from the runs discussed in the 
section below.
In the first simulation we used the leapfrog algorithm with $m=32$. From 
500 trajectories we get $P_{acc} = 0.874(8)$.  Note that from table I of
ref. \cite{lippert} we read off that without UV-filtering, $m=100$ results
in $P_{acc} = 0.78$. Hence we see a speed-up by roughly a factor of three, 
as it is the case for $\kappa=0.156$ and the $12^3 \times 24$ lattice.
In the second simulation we used the second order Omelyan scheme with 
$\lambda=1/6$ and $m=16$. Performing 500 trajectories we find 
$P_{acc} = 0.926(6)$. 

\subsection{Runs with rooted polynomials}

\begin{figure}
\begin{center}
\includegraphics[width=14.5cm]{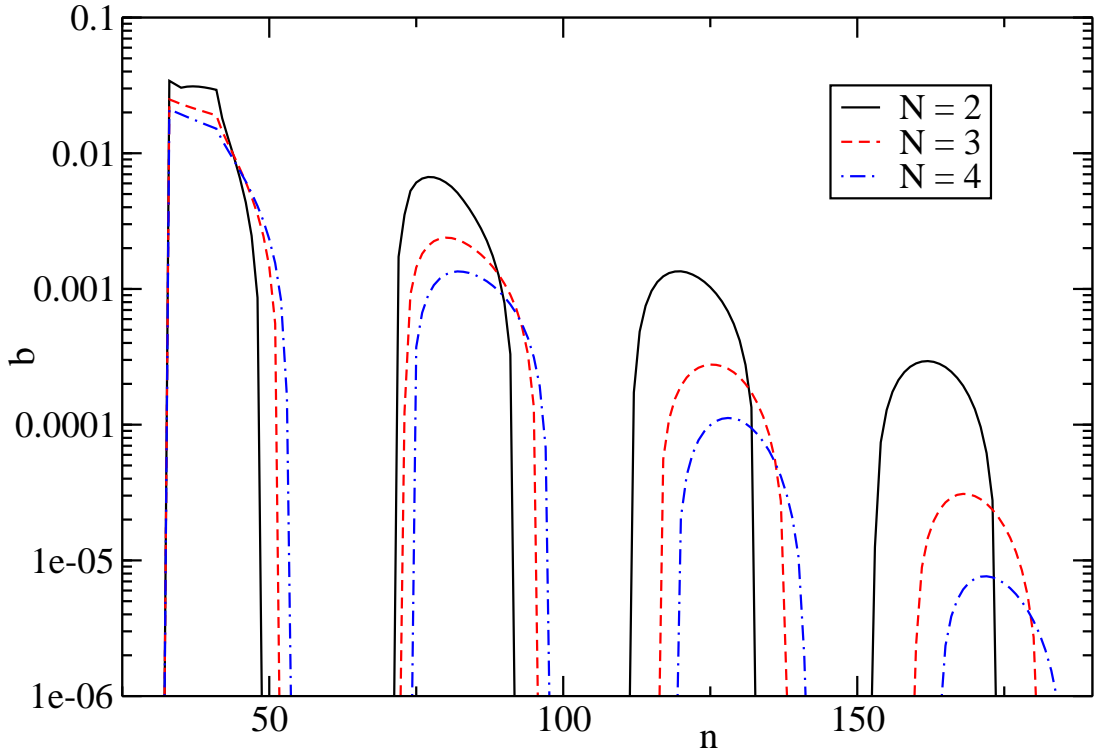}
\caption{\label{coefficients}  We give the coefficients $b_n$, 
eq.~(\ref{restM}), as a function of the index $n$ for $N=2, 3$ and $4$. 
Only results for $b_n \ge 10^{-6}$ are shown.
}
\end{center}
\end{figure}

We performed a few runs  with the rooted polynomial action.
We simulated a $16^3 \times 32$ 
lattice at $\beta=5.6$ and $\kappa=0.1575$.  All runs are characterized by 
$j_{max} =2$.  In all cases we use for simplicity the leapfrog scheme 
with different time scales.  
Throughout we use the trajectory length $\tau=\sqrt{2}$.

\subsubsection{Without hopping parameter expansion}
In this first set of runs we simulated without making use of the 
hopping parameter expansion. The polynomials are characterized by 
$n_1=8$ and $n_2=32$ and rooting with $N=2,3,4,6,8$, and $16$. 
In figure \ref{coefficients} we show the coefficient $b_n$ of eq.~(\ref{restM}) 
for $N=2,3$, and $4$.   For $n > n_2=32$,  $b_n$ is oscillating, with a  
decreasing amplitude. As it can be seen from the figure, the decay is 
exponential in $n$.  The decay becomes faster with increasing $N$.  
With increasing $N$, the decay rate converges to a finite limit.  
In table
\ref{RunPoly1} we summarize the basic parameters of the simulations and
give the acceptance rate $P_{acc}$ and Var$(\Delta H)$.  We have taken
$m$ such that $P_{acc} \approx 0.8$.  The  
parameters $m_2$, $m_1$, and $m_0$ are chosen ad hoc and are likely 
larger than the optimal values.
Note that error bars might be underestimated, since the lengths of the
runs are relatively short. It is reassuring
that our estimates of $\langle P \rangle$  
are consistent with the result given in table I of \cite{lippert}. 

\begin{table}
\caption{\sl \label{RunPoly1}
Basic properties of our runs with rooted polynomials. The polynomials 
approximate the $N^{th}$ root of $M^{-1}$. stat gives the number
of trajectories, $m_0$, ..., $m_2$ and $m$ 
are the number of steps that characterise
the multi-level integration scheme, and $\langle P \rangle$  is the expectation 
value of the plaquette. In the seventh column we give $\alpha$ of 
eq.~(\ref{restM}). Note that in the limit $N \rightarrow \infty$ we get 
$\alpha=0.017275...\;$. Finally, in the eighth and ninth column we give 
the acceptance rate $P_{acc}$ and the variance Var$(\Delta H)$, respectively.  
}
\begin{center}
\begin{tabular}{rccccccccc}
\hline
 $N$ & stat  & $m_0$ &  $m_1$  & $m_2$ &   $m$  & $\alpha$ & $\langle P \rangle$ & 
 $P_{acc}$ &Var$(\Delta H)$ \\
\hline
 2 & 290 & 6 &  6 & 4 & 8 & 0.022110... &0.57279(6)&0.870(11)& 0.110(9) \\
 2 & 510 & 4 &  4 & 2 & 8 & 0.022110... &0.57257(5)&0.836(12) & 0.191(16)  \\
 3 & 500 &10 &  6 & 3 & 6 & 0.020504... &0.57255(6)&0.788(9)  & 0.253(22)\\
 4 & 910 & 6 &  5 & 4 & 5 & 0.019687... &0.57258(3)&0.793(9)  & 0.299(16) \\
 6 & 600 &10 &  5 & 2 & 5 & 0.018872... &0.57256(5)&0.773(11) & 0.309(18) \\
 8 & 500 & 6 &  5 & 2 & 5 & 0.018467... &0.57254(5)&0.792(12) & 0.311(19) \\
16 & 200 & 6 &  5 & 2 & 5 & 0.017866... &0.57259(6)&0.806(17) & 0.241(25) \\
\hline
\end{tabular}
\end{center}
\end{table}
In table \ref{RunPolyForce} we summarize the results obtained for the 
variances of the forces.
\begin{table}
\caption{\sl \label{RunPolyForce}
Results for the variance of the forces  Var$(|{\cal F}_G|^2)$, 
Var$(|{\cal F}_{PF,1}|^2)$,  Var$(|{\cal F}_{PF,2}|^2)$, and 
Var$(|{\cal F}_{PF,3}|^2)$ from our simulations with rooted polynomials
listed in table \ref{RunPoly1}.
The fermion matrix is taken to the $N^{th}$ root.
}
\begin{center}
\begin{tabular}{ccccc}
\hline
  $N$ &  $G$  &  ${PF,1}$    &    ${PF,2}$    &  ${PF,3}$ \\
\hline
   2  & 85000000(4500000) &1110000(60000)&11000(900)& 1300(120) \\
   3  & 82000000(4700000) &290000(15000)&2020(180)  & 540(60) \\
   4  & 84000000(3500000) &114000(4000) &710(100)   & 360(60) \\  
   6  & 77000000(4000000) & 42400(2000) &197(17)    & 156(15) \\
   8  & 79000000(4000000) & 20200(1000) & 81(8)     & 123(14) \\ 
  16  & 83000000(6000000) &  4860(400)  & 16.5(3.0) & 156(40) \\ 
\hline
\end{tabular}
\end{center}
\end{table}
As one might expect, Var$(|{\cal F}_G|^2)$  does not depend on $N$. 
Furthermore, comparing with the runs for the $12^3 \times 24$ lattice 
of the previous section,  we see that  Var$(|{\cal F}_G|^2)$  is roughly 
proportional to the volume of the lattice. In the case of the rooted 
pseudo-fermion action we find that 
Var$(|{\cal F}_{PF,1}|^2)$ and Var$(|{\cal F}_{PF,2}|^2)$ are decreasing 
with increasing $N$. In the limit $N \rightarrow \infty$, a finite value, 
corresponding to the hopping parameter expansion should be reached. Here, 
it seems that we are still far away from this limit. Going from 
$N=8$ to $16$,  Var$(|{\cal F}_{PF,1}|^2)$ and Var$(|{\cal F}_{PF,2}|^2)$ 
are reduced by roughly a factor of four. Following  eq.~(\ref{masterF}), this 
should allow to increase the corresponding step-size by a factor of $\sqrt{2}$. 
Since the numerical effort for evaluating $S_{PF}$ 
 increases by a factor of two, the algorithm 
becomes less efficient. In order to compare the numerical costs, we 
define the cost index $c= n_j N $ Var$(|{\cal F}_{PF,1}|^2)^{1/4}$, where 
the exponent $1/4$ is motivated by eq.~(\ref{masterF}). Our results are
summarized in table \ref{Costf}. In the case of $S_{PF,1}$ we see a  small
increase from $N=2$ to $4$. For $S_{PF,2}$ the cost index is very similar
for $N=2, 3$, and $4$. On the other hand, Var$(|{\cal F}_{PF,3}|^2)$ is 
clearly decreasing going from $N=2$ to $4$. The costs related with $S_{PF,3}$
depend on the solver that is used. Here we made no effort to find the 
optimal solver. 
Therefore we refrain from quoting a performance index for $S_{PF,3}$. 
Anyway, it seems likely that  the optimal overall performance is reached
for $N > 2$. 

\begin{table}
\caption{\sl \label{Costf}
Cost index related to the terms $S_{PF,1}$ and 
$S_{PF,2}$ of the pseudo-fermion action. The estimates of Var$(|{\cal F}|^2)$
are taken from table \ref{RunPolyForce}. 
}
\begin{center}
\begin{tabular}{ccc}
\hline
  $N$ & $8 N$ Var$(|{\cal F}_{PF,1}|^2)^{1/4}$ & 
$32 N$ Var$(|{\cal F}_{PF,2}|^2)^{1/4}$ \\
\hline
    2 & 519(7) &   655(13) \\ 
    3 & 557(7) &   644(14) \\ 
    4 & 588(5) &   661(22) \\ 
    6 & 689(8) &   719(15) \\
    8 & 763(9) &   768(18) \\  
   16 &1069(21)&  1032(44) \\  
\hline
\end{tabular}
\end{center}
\end{table}

\subsubsection{Employing $\kappa^4$-filtering}
UV-filtering by using the hopping parameter expansion can by easily 
implemented in the PHMC-algorithm \cite{Ishikawa:2006pb}. Here we 
perform a preliminary study, employing $\kappa^4$-filtering. We consider
polynomials characterized by $n_1=16$ and $n_2=42$ and  $N=8$.
The remainder is characterized by $\alpha=0.01390254...\;$. Note that in the 
limit $N \rightarrow \infty$ one gets $\alpha= 0.01321050...\;$.
The parameters of the HMC are 
$m=4$, $m_{2}=2$, $m_{1}=3$, $m_{0}=40$, and $n_t=160$. 
We performed 500 trajectories.   
The acceptance rate is $P_{acc}=0.790(10)$ and Var$(\Delta H)=0.249(22)$.
For the variances of the 
forces we get Var$(|{\cal F}_G|^2) = 90000000(5000000)$, 
Var$(|{\cal F}_{PF,1}|^2) = 2370(160)$, Var$(|{\cal F}_{PF,2}|^2) = 23.8(2.5)$
and  Var$(|{\cal F}_{PF,3}|^2) = 42(5)$.  In particular 
Var$(|{\cal F}_{PF,1}|^2)$ is considerably reduced compared with the 
run for $N=8$, discussed above. The result $\langle P \rangle= 0.57265(5)$
for the plaquette is consistent with that given in table I of \cite{lippert}. 

\section{Conclusion and outlook}
We discuss how the hopping parameter expansion can be used as an efficient
UV-filter in the HMC simulation of lattice QCD with two degenerate fermion 
flavours. 
We have carefully tested the idea for the Wilson gauge action 
and Wilson fermions at $\beta=5.6$ and $\kappa=0.156$ and the 
relatively small lattice size $12^3 \times 24$. Compared with the 
pseudo-fermion action~(\ref{WePe_action}) we find a speed-up of a factor of 
two and three, using  $\kappa^2$- and $\kappa^4$-filtering, 
respectively. The latter result is confirmed by short runs performed for
a $16^3 \times 32$ lattice and $\kappa=0.1575$.

In large scale simulations the idea can be combined with  mass 
preconditioning or domain decompositioning. In the case of 
mass preconditioning one might be able to skip the term in the action that
corresponds to the most heavy mass. In the case of domain decompositioning
one applies the idea to the fermion matrix that is restricted  to the 
domains. The speed-up achieved this way  might be of the order of $20 \%$.

A natural extension of applying the hopping parameter expansion as UV-filter
is the use of rooted polynomials. This idea is related with the rooting 
proposed in ref. \cite{Clark007} as well as the idea of hierarchically 
factorised polynomials \cite{Kamleh17,Kamleh18}. Here our results are still
preliminary, and both a better theoretical understanding as well as further 
numerical experiments are needed.

\section{Acknowledgement}
This work was supported by the Deutsche Forschungsgemeinschaft under the 
grant No HA 3150/4-1.


\begin{thebibliography}{99}
\bibitem{Rothe}
H.J. Rothe, 
{\sl Lattice Gauge Theories: An Introduction}, 
World Sci. Lect. Notes Phys. {\bf 43},1 (1992),  {\bf 82}, 1 (2012).

\bibitem{review}
I. Montvay and G. M\"unster, {\sl Quantum fields on a lattice}, (Cambridge 
University Press, Cambridge, 1994). 

\bibitem{review2}
C. Gattringer and C. B. Lang, 
{\sl Quantum Chromodynamics on the Lattice: An Introductory Presentation}, 
Lect. Notes Phys. {\bf 788} (Springer, Berlin Heidelberg, 2010) 

\bibitem{Gupta97}
R. Gupta,
{\sl Introduction to Lattice QCD}, 
Lectures given at the LXVIII Les Houches Summer School 
"Probing the Standard Model of Particle Interactions", [ arXiv:hep-lat/9807028] 

\bibitem{WePe}
D. Weingarten and D. Petcher, 
{\sl  	
Monte Carlo Integration for Lattice Gauge Theories with Fermions}, 
Phys.\ Lett.\ B {\bf 99}, 333 (1981).

\bibitem{HMC}
S. Duane,  A. D. Kennedy, B. J. Pendleton, and D. Roweth,
{\sl Hybrid Monte Carlo},
Phys.\ Lett.\ B {\bf 195}, 216 (1987).

\bibitem{domain2}
M. L\"uscher,
{\sl Schwarz-preconditioned HMC algorithm for two-flavour lattice QCD},
[arXiv:hep-lat/0409106],
Comput.\ Phys.\ Commun.\ {\bf 165}, 199 (2005).

\bibitem{Gupta89}
R. Gupta, A. Patel, C.F. Baillie, G. Guralnik, G.W. Kilcup, and S.R. Sharpe,
{\sl  	
QCD With Dynamical Wilson Fermions}, 
Phys.\ Rev.\ D {\bf 40}, 2072 (1989).

\bibitem{Schaefer}
S. Schaefer, R. Sommer, and F. Virotta, 
{\sl Critical slowing down and error analysis in lattice QCD},
[arXiv:1009.5228], Nucl.\ Phys.\ B {\bf 845}, 93  (2011). 

\bibitem{phialg}
S.A. Gottlieb, W. Liu, D. Toussaint,  R.L. Renken, and R.L. Sugar,
{\sl Hybrid Molecular Dynamics Algorithms for the Numerical Simulation of 
Quantum Chromodynamics},
Phys.\ Rev.\ D {\bf 35}, 2531 (1987).

\bibitem{Omel03}
I.P. Omelyan, I.M. Mryglod, and R. Folk, 
{\sl Symplectic analytically integrable decomposition algorithms: 
Classification, derivation, and application to molecular dynamics, 
quantum and celestial mechanics simulations}, 
Comput.\ Phys.\ Commun. 
{\bf 151}, 272 (2003).

\bibitem{Sex92}
J.C. Sexton and D.H. Weingarten, 
{\sl	
Hamiltonian evolution for the hybrid Monte Carlo algorithm}, 
Nucl.\ Phys.\ B {\bf 380}, 665 (1992).

\bibitem{Bussone18}
A. Bussone, M. Della Morte, V. Drach, and C. Pica,
{\sl Tuning the Hybrid Monte Carlo algorithm using molecular dynamics forces' 
variances}, [arXiv:1801.06412].

\bibitem{urbach}
C. Urbach, K. Jansen, A. Shindler, and U. Wenger,
{\sl HMC algorithm with multiple time scale integration and mass 
preconditioning}, [arXiv:hep-lat/0506011], 
Comput.\ Phys.\ Commun.\ {\bf174}, 87 (2006).

\bibitem{shadow}
A.D. Kennedy, P.J. Silva, and M.A. Clark, 
{\sl Shadow Hamiltonians, Poisson Brackets, and Gauge Theories},
[arXiv:1210.6600], Phys.\ Rev.\ D 87, 034511 (2013).

\bibitem{BiCG1}
H. A. Van der Vorst, {\sl Bi-CGSTAB: A Fast and Smoothly Converging Variant of 
Bi-CG for the Solution of Nonsymmetric Linear Systems},
SIAM J.\ Sci.\ and Stat.\ Comput.\ {\bf 13-2}, 631 (1992).

\bibitem{BiCG2}
M. H. Gutknecht, {\sl Variants of BICGSTAB for Matrices with Complex Spectrum}, SIAM J.\ Sci.\ Comput.\ {\bf 14-5}, 1020 (1993).

\bibitem{Luescherdefl}
M. L\"uscher, 
{\sl Local coherence and deflation of the low quark modes in lattice QCD},
[arXiv:0706.2298],  J. High Energy Phys. {\bf 07} (2007) 081. 

\bibitem{Multigrid}
 R. Babich
et al.,
{\sl Adaptive multigrid algorithm for the lattice Wilson-Dirac operator},
[arXiv:1005.3043], Phys.\ Rev.\ Lett.\ {\bf 105}, 201602, (2010).

\bibitem{multiboson}
M. L\"uscher,
{\sl A New Approach to the Problem of Dynamical Quarks in Numerical
Simulations of  Lattice QCD},
[arXiv:hep-lat/9311007], Nucl.\ Phys.\ B {\bf 418}, 637 (1994).

\bibitem{Myfinite}
 M. Hasenbusch,
{\sl Speeding up finite step-size updating of full QCD on the lattice},
[arXiv:hep-lat/9807031], Phys.\ Rev.\ D {\bf 59}, 054505 (1999).

\bibitem{Knechtli13}
J. Finkenrath, F. Knechtli, and B. Leder, 
{\sl Fermions as Global Correction: the QCD Case}, 
[arXiv:1204.1306], Comput.\ Phys.\ Commun.\ {\bf 184}, 1522 (2013).

\bibitem{MyHasenbusch}
M. Hasenbusch,
{\sl Speeding up the hybrid Monte Carlo algorithm for dynamical fermions},
[arXiv:hep-lat/0107019], Phys.\ Lett.\ B {\bf 519}, 177 (2001). 

\bibitem{masspreconditioning2}
M. Hasenbusch and K. Jansen,
{\sl Speeding up Lattice QCD simulations with clover-improved Wilson Fermions},
[arXiv:hep-lat/0211042], Nucl.\ Phys.\ B {\bf 659}, 299 (2003).

\bibitem{domain}
M. L\"uscher, {\sl Lattice QCD and the Schwarz alternating procedure},
[arXiv:hep-lat/0304007], J. High Energy Phys. {\bf 05} (2003) 052. 

\bibitem{Clark007}
M. A. Clark and A. D. Kennedy,
{\sl Accelerating dynamical fermion computations using the rational hybrid
 Monte Carlo (RHMC) algorithm with multiple pseudo-fermion fields},
[arXiv:hep-lat/0608015], Phys.\ Rev.\ Lett.\ {\bf 98}, 051601 (2007).

\bibitem{ForcrandPHMC}
Ph. de Forcrand and T. Takaishi, 
{\sl  	
Fast fermion Monte Carlo}, [arXiv:hep-lat/9608093],
Nucl.\ Phys.\ Proc.\ Suppl.\  {\bf 53}, 968 (1997).

\bibitem{PHMC}
R. Frezzotti and K. Jansen, 	
{\sl A Polynomial hybrid Monte Carlo algorithm}, 
[arXiv:hep-lat/9702016], 
Phys.\ Lett.\ B {\bf 402}, 328 (1997).

\bibitem{PHMC2}
S. Aoki et al., {Polynomial hybrid Monte Carlo algorithm for lattice
QCD with an odd number of flavors}, [arXiv:hep-lat/0112051],
Phys.\ Rev.\ D {\bf 65}, 094507 (2002).

\bibitem{ForcrandUV} 
Ph. de Forcrand,
{\sl 
UV filtered fermionic Monte Carlo}, 
[arXiv:hep-lat/9809145], Nucl.\ Phys.\ Proc.\ Suppl.\ {\bf 73}, 822 (1999).

\bibitem{Forcrand}
C. Alexandrou, Ph. de Forcrand, M. D'Elia, and H. Panagopoulos,
{\sl Efficiency of the UV-filtered Multiboson algorithm}, 
[arXiv:hep-lat/9906029], Phys.\ Rev.\ D {\bf 61}, 074503 (2000).

\bibitem{Ishikawa:2006pb} 
  K.-I. Ishikawa {\it et al.} [PACS-CS Collaboration],
  {\sl An Application of the UV-filtering preconditioner to the polynomial 
  hybrid Monte Carlo algorithm}, [arXiv:hep-lat/0610037]
  PoS LAT {\bf 2006}, 027 (2006).

\bibitem{thron97}
C. Thron, S.J. Dong, K.F. Liu, and H.P. Ying,
{\sl Pad\'e-$Z_2$ estimator of determinants}, 
[arXiv:hep-lat/9707001], Phys.\ Rev.\  D {\bf 57}, 1642 (1998).



\bibitem{Kamleh12}
W. Kamleh and M. Peardon,
{\sl Polynomial Filtered HMC: 
An Algorithm for lattice QCD with dynamical quarks}, 
[arXiv:1106.5625], Comput.\ Phys.\ Commun.\ {\bf 183}, 1993 (2012).

\bibitem{Kamleh17}
T. Haar, W. Kamleh, J. Zanotti, and Y. Nakamura, 
{\sl Improving Polynomial-filtered Hybrid Monte Carlo with Hasenbusch},
[arXiv:1702.00124], PoS {\bf INPC2016}, 319 (2017).

\bibitem{Kamleh18}
W. Kamleh, T. Haar, Y. Nakamura, and J. Zanotti,
{\sl Single flavour filtering for RHMC in BQCD},
[arXiv:1711.07167],  EPJ Web Conf.\ {\bf 175}, 09004 (2018).
 	
\bibitem{Forcrand99}
Ph. de Forcrand, 
{Monte Carlo quasi-heat-bath by approximate inversion}, 
[arXiv:cond-mat/9811025], Phys.\ Rev.\ E {\bf 59}, 3698 (1999).

\bibitem{Thesis}
B. Stra{\ss}berger,  {\sl Evaluation of Fermion Determinant Splitting in 
Flavor Lattice QCD Simulations}, Master thesis, Humboldt-Universit\"at zu 
Berlin (2017). 





\bibitem{lippert}
B. Orth, T. Lippert, and K. Schilling, 
{\sl Finite-Size Effects in Lattice QCD with Dynamical Wilson Fermions},
[arXiv:hep-lat/0503016], Phys.\ Rev.\ D {\bf 72}, 014503 (2005).

\bibitem{Sommer} 
R. Sommer, 
{\sl A New Way to Set the Energy Scale in Lattice Gauge Theories and its 
Application to the Static Force and $\alpha$s in SU(2) Yang--Mills Theory},
[hep-lat/9310022], Nucl.\  Phys.\ B {\bf 411}, 839 (1994).





\end{thebibliography}
\end{document}